\documentclass[aps,prl,preprint,superscriptaddress]{revtex4-1}

\usepackage{amsmath}
\usepackage{cases}
\usepackage{graphicx}
\usepackage{rotating}
\usepackage{color}  
\usepackage{enumitem}
\usepackage{multirow}
\usepackage{tabularx}
\usepackage{longtable}
\usepackage{float}

\usepackage{hyperref}
\usepackage{trkchg}

\begin{document}


\title{Approximate Green's Function Coupled Cluster Method Employing Effective Dimension Reduction}

\author{Bo Peng} 
\email{peng398@pnnl.gov}              
\affiliation{William R. Wiley Environmental Molecular Sciences Laboratory, Battelle, 
              Pacific Northwest National Laboratory, K8-91, P.O. Box 999, Richland, WA 99352, USA}
\author{Roel Van Beeumen}
\email{rvanbeeumen@lbl.gov}
\author{David B. Williams-Young}
\email{dbwy@lbl.gov}
\affiliation{Computational Research Division, Lawrence Berkeley National Laboratory, Berkeley, CA 94720, USA}
\author{Karol Kowalski} 
\email{karol.kowalski@pnnl.gov}              
\affiliation{William R. Wiley Environmental Molecular Sciences Laboratory, Battelle, 
              Pacific Northwest National Laboratory, K8-91, P.O. Box 999, Richland, WA 99352, USA}
\author{Chao Yang}
\email{cyang@lbl.gov}
\affiliation{Computational Research Division, Lawrence Berkeley National Laboratory, Berkeley, CA 94720, USA}


\date{\today}

\begin{abstract}
The Green's function coupled cluster (GFCC) method, originally proposed in the early 1990s, is a powerful many-body tool for computing and analyzing the electronic structure of molecular and periodic systems, especially when electrons of the system are strongly correlated.
However, in order for the GFCC to become a method that may be routinely used in the electronic structure calculations, more robust numerical techniques and approximations must be employed to reduce its extremely high computational overhead.
In our recent studies, it has been
demonstrated that the GFCC equations can be solved directly in the frequency domain using
iterative linear solvers, which can be easily distributed in a massively parallel environment.
In the present work, we demonstrate a successful application of model-order-reduction (MOR) techniques in the GFCC framework. Briefly speaking, for a frequency regime of interest which requires high resolution descriptions of spectral function, instead of solving GFCC linear equation of full dimension for every single frequency point of interest, an efficiently-solvable linear system model of a reduced dimension may be built upon projecting the original GFCC linear system onto a subspace. From this reduced order model is obtained a
reasonable approximation to the full dimensional GFCC linear equations in both interpolative and
extrapolative spectral regions. Here, we show that the subspace can be properly constructed in an iterative manner from the auxiliary vectors of the GFCC linear equations at some selected frequencies within the spectral region of interest. 
During the iterations, the quality of the subspace, as well as the linear system model, can be systematically improved. The method is tested in this work in terms of the efficiency and accuracy of computing spectral functions for some typical molecular systems such as carbon monoxide, 1,3-butadiene, benzene, and adenine. 
As a byproduct, the reduced order model obtained by this method is also found to provide a high quality initial guess which improves the convergence rate for the existing iterative linear solver.
\end{abstract}


\maketitle

\section{Introduction}

Green's function formalisms\cite{paldus74_149,paldus75_105,mattuck2012guide,abrikosov2012methods,fetter2012quantum,migdal1967theory} are a powerful tool for describing excitation and correlation phenomena associated with molecules, clusters, nano-structures, and solids. Numerous approximations and methods in the Green's function framework have been developed over the years. Typical developments include
the outer valence Green's function (OVGF) method\cite{cederbaum75_290, cederbaum84_57, ortiz97, ortiz13_123}, 
the diagonal Green's function approximations\cite{zakrzewski2010ab},
the non-diagonal renormalized second-order method\cite{ortiz98_1008},
the generalized perturbative methods\cite{hirata15_1595, hirata17_044108},
the two-hole-one-particle Tamm-Dancoff  approximation (2h-p TDA)\cite{cederbaum77_4124},
the third-order quasiparticle method\cite{ortiz96_7599},
the algebraic-diagrammatic construction (ADC) methods\cite{schirmer82_2395,cederbaum84_57, dreuw15_82}, 
the $GW$-related methods\cite{hedin65_a796,rehr00_621,schilfgaarde06_226402,louie11_186404,kas14_085112,reining18_e1344},
and the Green's function coupled cluster (GFCC) method\cite{nooijen92_55, nooijen93_15, nooijen95_1681}.
There are also efforts motivated by the development of various embedding methods, such as dynamical mean field theories (DMFT)\cite{kotliar96_13, kotliar06_865, vollhardt12_1, millis06_155107, millis06_076405, zgid11_094115, zgid12_165128} and 
self-energy embedding theory (SEET)\cite{zgid15_121111,zgid15_241102,zgid16_054106,zgid17_2200}, to develop efficient ways to include higher-order correlation in the Green's function calculation of larger systems.
Many applications of these methods have been reported in the literature of several research areas of computational physics, chemistry, and materials\cite{onida2002_601,faleev04_126406,louie06_216405}. 

In scenarios where the accurate description of many-body effects is required, only a limited subset of the aforementioned developments may
be reliably employed.
%
One such choice which has drawn some attention
recently is the GFCC method. In essence, the GFCC method
belongs to a larger class of methods based on intermediate-state representations (ISRs)\cite{mertins96_2140,mertins96_2153}.
Similar to the linear response coupled cluster (LRCC)\cite{monkhorst77_421,jorgensen90_3333} and equation-of-motion coupled cluster (EOM-CC)\cite{geertsen89_57,comeau93_414,stanton93_7029} methods, the GFCC method employs the bi-orthogonal formalism\cite{Schirmer10_145, helgaker2014molecular} of the coupled cluster (CC) theory to express Green's function matrix in terms of the cluster operator and the so-called $\Lambda$ operator. 
Practically, there are two primary methods
to numerically evaluate the CC Green's function.
The most straightforward way is to directly diagonalize the non-Hermitian EOM-CC Hamiltonian matrix in the ($N\pm1$)-particle space (i.e. ionization potential and electron attachment EOM-CC method, or IP/EA-EOM-CC) to obtain all the eigenpairs needed to construct the sum-over-states representation of the CC Green's function (see Ref. \citenum{kowalski16_144101} for more details). 
As the dimension of the EOM-CC Hamiltonian matrix grows polynomially with
the number of orbitals used to describe the system, direct diagonalization via methods such
as those provided by LAPACK and ScaLAPACK quickly become infeasible. If the spectral
region of interest is in the extreme ends of the spectrum of the EOM-CC Hamiltonian,
one may utilize standard iterative Krylov subspace methods, such as Arnoldi's method and
variants\cite{arno1951,bade2000}, to obtain the eigenpairs of interest. The power of methods based on
Krylov subspace iterations is in that they are matrix-free, i.e. one need only knowledge of
the matrix-vector product to perform the partial diagonalization. However, if the spectral
region of interest is in the spectral interior of the EOM-CC Hamiltonian,
the standard Arnoldi
method exhibits poor convergence for these embedded states,
and one should employ spectral transformations to achieve proper convergence behavior\cite{bade2000}.

A typical workaround to obtain these embedded states is to introduce approximations such as core-valence separation (CVS)\cite{cederbaum80_206}. 
In the CVS, one neglects the coupling between core- and valence-excited states such that the dimension of the effective Hamiltonian for the problem of interest may be reduced and the convergence to the target states may be accelerated. There have been studies reporting  the success of the CVS approximation in relation to the ADC, CC2, CCSD, CC3, and CCSDR(3) methods (see Ref. \citenum{norman18_7208} for a recent review). For example, in the ADC calculations of the K-shell ionization spectra of small and medium-size molecules\cite{schirmer87_6789, dreuw14_4583, trofimov00_483}, main ionic states, as well as some satellite states, may be efficiently obtained with reasonable accuracy (0.4$\sim$1.0 eV) when the CVS is employed. However, if smaller errors and higher resolution of the spectra are required for the system of
interest, the use of CVS may not be sufficient. 
In the context of the EOM-CC calculations, recent years have also
seen several developments in the robust application of Krylov subspace methods to obtain
embedded electronic states through methods such as
asymmetric Lanczos-chain-driven subspace algorithm\cite{coriani12_1616}, energy-specific Davidson algorithm\cite{peng15_4146}, and Generalized Preconditioned Locally Harmonic Residual (GPLHR) method\cite{zuev14_273}. 

Alternatively, 
the CC Green's function matrix
may be obtained directly in its analytic form\cite{nooijen92_55, nooijen93_15, nooijen95_1681} through the evaluation of a shifted set of
linear systems involving the IP/EA-EOM-CC Hamiltonian at a frequency of interest\cite{kowalski14_094102, kowalski16_144101,kowalski16_062512, kowalski18_561,kowalski18_4335, kowalski18_214102}. 
In the treatment of the GFCC method as a linear system, one is able to bypass the need
for the evaluation of the eigenstates of the IP/EA-EOM-CC Hamiltonian explicitly, and in principle efficiently resolve the entire spectrum of the IP/EA-EOM-CC Hamiltonian. Furthermore, due to its algebraic structure, the solution of the linear systems posed by the GFCC method
may be executed in a highly scalable manner which is well suited for massively distributed
computing architectures\cite{kowalski18_4335}. Using a similar method for solving the GFCC linear systems as employed in this work, there have been studies recently reported for the computation of the spectral function of, for example, uniform electron gas,\cite{chan16_235139} light atoms,\cite{matsushita18_034106} heavy metal atoms,\cite{matsushita18_224103} and simple 1-D periodic systems.\cite{matsushita18_204109}
In addition, the application of the GFCC method to more
complicated 3-D periodic systems seems imminent in that the relevant EOM-CC formalism
has been recently developed\cite{chan16_235139,chan17_1209}. 

Nevertheless, solving the GFCC linear system may be computationally demanding. In particular, if a much higher frequency resolution is necessary to explore some detailed information (for example the state of interest with weak intensity) or a much broad frequency domain needs to be explored, the number of frequency points will become a very large computational overhead of the GFCC method which prohibits its larger scale applications. 
Fortunately, it seems there exists some interpolation techniques that can be utilized to cleverly address this problem. One such choice is the model-order-reduction (MOR) technique. The MOR technique is originally introduced in the field of systems and control to reduce the computational complexity of mathematical models in
numerical simulations. The main motivation for using MOR techniques comes from limited
computational, accuracy, or storage capabilities. After applying MOR, the reduced and
simplified model, which still captures the main features of the original large model, can be
used in place of the original one at a much lower cost (see Ref. \citenum{Antoulas2005} for an overview of MOR techniques for linear dynamical systems). 


In the present study, we examine the potential to reduce the
computational overhead of the GFCC method by applying MOR techniques
to interpolate and extrapolate the evaluation of the CC Green's function. This work has
been inspired, in part, by the recent application of MOR techniques by Van Beeumen, et
al.\cite{vanbeeumen17_4950} to obtain
accurate approximations of the X-Ray absorption spectrum within the time-dependent
Hartree-Fock approximation for water clusters with reduced computational complexity.
In this work, we project
the full dimensional GFCC linear system onto a subspace which serves as approximation
to exact CC Green's function for a broad spectral range of interest.
The quality and efficiency of the proposed method for a series of
representative molecular systems, such as carbon monoxide, 1,3-butadiene, benzene, and adenine,
will be assessed in terms of the reduction of computational overhead and its comparison to
other theoretical and experimental results.


\section{Methodology}

For a brief review of the GFCC method employed in this work, we refer the readers to Refs. \citenum{kowalski14_094102, kowalski16_144101,kowalski16_062512, kowalski18_561,kowalski18_4335, kowalski18_214102}. For the purposes of this work, it is
sufficient to consider the matrix element of the retarded part of the analytical frequency dependent CC Green's function of an $N$-electron system, 
\begin{eqnarray}
G^R_{pq}(\omega) =
&&\langle \Psi | a_q^\dagger (\omega + ( H - E_0 ) - i \eta)^{-1} a_p | \Psi \rangle
\label{gfxn0}
\end{eqnarray}
where $H$ is the electronic Hamiltonian of the $N$-electron system, $| \Psi \rangle$ is the normalized ground-state wave function of the system, $E_0$ is the ground state energy, and the $a_p$ ($a_p^\dagger$) operator is the annihilation (creation) operator for electron in the $p$-th spin-orbital. Besides, $\omega$ is the frequency, $\eta$ is the broadening factor, and $p,q,r,s,\ldots$ refers to general spin-orbital indices (we also use $i,j,k,l,\ldots$ to label occupied spin-orbital indices, and $a,b,c,d,\ldots$ to label virtual spin-orbital indices). By introducing bi-orthogonal CC formalism, the CC Green's function can then be expressed as
\begin{eqnarray}
G^R_{pq}(\omega) = 
&&\langle\Phi|(1+\Lambda) \bar{a_q^{\dagger}} (\omega+\bar{H}_N- \text{i} \eta)^{-1} 
	\bar{a}_p |\Phi\rangle 
\label{gfxn1}
\end{eqnarray}
where $|\Phi\rangle$ is the reference function, and the normal product form of similarity transformed Hamiltonian $\bar{H}_N$ is defined as $\bar{H} - E_0$. The similarity transformed operators $\bar{A}$ ($A = H, a_p, a_q^{\dagger}$) are defined as $\bar{A} = e^{-T} A ~e^{T}$. The cluster operator $T$ and the de-excitation operator $\Lambda$ are obtained from solving the conventional CC equations. Now we can introduce an $\omega$-dependent IP-EOM-CC type operators $X_p(\omega)$ mapping the $N$-electron Hilbert space onto an ($N$$-$1)-electron Hilbert space
\begin{eqnarray}
X_p(\omega) 
&=& X_{p,1}(\omega)+X_{p,2}(\omega) + \ldots \notag \\
&=& \sum_{i} x^i(p, \omega)  a_i  + \sum_{i<j,a} x^{ij}_a(p, \omega) a_a^{\dagger} a_j a_i +\ldots ~, \label{xp} 
\end{eqnarray}
that satisfies 
\begin{eqnarray}
(\omega+\bar{H}_N - \text{i} \eta )X_p(\omega)|\Phi\rangle = 
	\bar{a}_p |\Phi\rangle.  \label{eq:xplin} 
\end{eqnarray}
Substituting this expression into Eq. (\ref{gfxn1}), we end up with a compact expression for the matrix element of the retarded CC Green's function
\begin{eqnarray}
G^R_{pq}(\omega) = 
\langle\Phi|(1+\Lambda) \bar{a_q^{\dagger}} X_p(\omega) |\Phi\rangle, 
\label{gfxn2}
\end{eqnarray}
which becomes
\begin{eqnarray}
G_{pq}(\omega) &=&  
\langle\Phi|(1+\Lambda_1+\Lambda_2) \bar{a_q^{\dagger}} (X_{p,1}(\omega)+X_{p,2}(\omega)) |\Phi\rangle 
\label{gfxn2}
\end{eqnarray}
in the GFCCSD approximation (GFCC with singles and doubles) with $X_{p,1}$/$\Lambda_1$ and $X_{p,2}$/$\Lambda_2$ being one- and two-body component of $X_{p}$/$\Lambda$ operators, respectively. The spectral function is then given by the trace of the imaginary part of the retarded GFCCSD matrix,
\begin{equation}
A(\omega) = - \frac {1} {\pi} \text{Tr} \left[ \Im\left({\bf G}^{\text{R}}(\omega) \right) \right] 
= - \frac {1} {\pi} \sum_{p} \Im\left(G_{pp}^{\text{R}}(\omega) \right)~.
\end{equation}

As discusseded in our previous work on this subject\cite{kowalski14_094102, kowalski16_144101,kowalski16_062512, kowalski18_561,kowalski18_4335, kowalski18_214102}, the practical calculation of GFCCSD matrix employing the above method involves the solution of the conventional CCSD calculations (to get converged $T$ and $\Lambda$ cluster amplitudes), solving linear systems of the form of Eq. (\ref{eq:xplin}) for all the orbitals ($p$'s) and frequencies of interest ($\omega$'s), and performing Eq. (\ref{gfxn2}). The key step is to solve Eq. (\ref{eq:xplin}) for 
$X_p(\omega)$ for given orbital $p$ and frequency $\omega$, and the overall computational cost approximately scales as $\mathcal{O}(N_{\omega}N^6)$ with the $N_{\omega}$ being the number of frequencies in the designated frequency regime. Therefore, if a finer or broader frequency range needs to be computed, $N_{\omega}$ would constitute a sizable pre-factor, and  dramatically increase the computational cost. 

To address this issue, in the context of high performance computing, one can divide the full computational task posed by the GFCC method into several smaller tasks according to the number of orbitals and frequencies desired.  In so doing, one can distribute these smaller tasks over the available processors to execute them concurrently. In this way, the overall computational cost remains the same, but the time-to-solution can be significantly reduced. In order to reduce the formal computational cost of the GFCC method, we must further introduce some approximations in our calculations from the perspective of the linear system solution method. Here, inspired by a similar application in the context
of the time-dependent Hartree-Fock method\cite{vanbeeumen17_4950}, we experiment the MOR technique in the context of the GFCC method. 



We can first represent Eqs. (\ref{eq:xplin}) and (\ref{gfxn2}) as a linear multiple-input multiple-output (MIMO) system $\mathbf{\Theta}$,
\begin{eqnarray}
\mathbf{\Theta}(\omega) = \left\{ \begin{array}{ccl}
(\omega - \text{i} \eta + \bar{\textbf{H}}_N) \textbf{X}(\omega) & = & \textbf{b}, \\
\mathbf{G}^{\text{R}}(\omega) & = & \textbf{c}^{\text{T}} \textbf{X}(\omega).
\end{array}\right. \label{gfcclineqn}
\end{eqnarray}
Here, the dimension of $\bar{\textbf{H}}_N$ is $D$ (For the GFCCSD method, $D$ scales as $\mathcal{O}$($N_o^2N_v$) with $N_o$ being the total number of occupied spin-orbitals and $N_v$ being the total number of virtual spin-orbitals). The columns of $\textbf{b}$ corresponding to free terms, and the rows of $\textbf{c}$ corresponding to $\langle \Phi | (1+\Lambda) \bar{a^\dagger_q}$. 
The transfer function of the linear system $\mathbf{\Theta}$,
\begin{eqnarray}
\mathbf{\gamma}(\omega) = \mathbf{c}^{\mathrm{T}} \left(\omega - \text{i} \eta + \bar{\textbf{H}}_N \right)^{-1} \mathbf{b}
\end{eqnarray}
describes the relation between the input and output of $\mathbf{\Theta}$, and is equal to its output $\mathbf{G}^{\mathrm{R}}(\omega)$ for \eqref{gfcclineqn}.

To apply the interpolation based MOR technique for $\mathbf{\Theta}$, we need to construct an orthonormal subspace
$\textbf{S} = \{ \textbf{v}_1, \textbf{v}_2, \cdots, \textbf{v}_m\}$ with $m \ll D$ and 
$\langle \textbf{v}_i | \textbf{v}_j \rangle  = \delta_{ij}$, such that the original linear system (\ref{gfcclineqn}) can be projected
into to form a model system $\hat{\mathbf{\Theta}}$,
\begin{eqnarray}
\hat{\mathbf{\Theta}}(\omega) = \left\{ \begin{array}{ccl}
(\omega - \text{i} \eta + \hat{\bar{\textbf{H}}}_N) \hat{\textbf{X}}(\omega) & = & \hat{\textbf{b}}, \\
\hat{\mathbf{G}}^{\text{R}}(\omega) & = & \hat{\mathbf{c}}^{\text{T}} \hat{\textbf{X}}(\omega),
\end{array}\right. \label{model}
\end{eqnarray}
where $\hat{\bar{\textbf{H}}}_N = \textbf{S}^{\text{T}} \bar{\textbf{H}}_N \textbf{S}$, $\hat{\textbf{X}}(\omega) = \textbf{S}^{\text{T}} \textbf{X}(\omega)$, $\hat{\textbf{b}} = \textbf{S}^{\text{T}} \textbf{b}$, and $\hat{\mathbf{c}}^{\text{T}} = \mathbf{c}^{\text{T}} \textbf{S}$. With the proper construction of the subspace $\textbf{S}$, we expect $\hat{\mathbf{\Theta}}(\omega) \approx \mathbf{\Theta}(\omega)$ for designated frequency regime.

In practice, the subspace $\textbf{S}$ is composed of the orthonormalized auxiliary vectors, $\textbf{X}_p$,  converged at selected frequencies $\omega_k$ in a given frequency regime, [$\omega_\text{min}$, $\omega_\text{max}$]. Hence, the transfer function of the reduced model $\hat{\mathbf{\Theta}}$ interpolates the original model $\mathbf{\Theta}$ at these selected frequencies, i.e.,
\begin{equation*}
\hat{\mathbf{G}}^\mathrm{R}(\omega_k) = \mathbf{G}^\mathrm{R}(\omega_k)
\end{equation*}
for $k = 1,\ldots,m$.
The sampling of the selected frequencies in the regime follows the adaptive refinement strategy described in Ref. \cite{vanbeeumen17_4950}. Basically, one can start with a uniformly sampled frequencies in the regime to construct a preliminary level reduced order model. 
Then, based on the error estimates of the computed spectral function between adjacent frequencies over the entire regime, one can decide whether the corresponding midpoints between these adjacent frequencies need to be added to refine the sampling. This refinement process continues until the maximal error estimate of the computed spectral function at the entire frequency window is below the threshold or when the refined model order exceeds a prescribed upper bound.

\begin{sidewaysfigure}
\includegraphics[width=\textwidth]{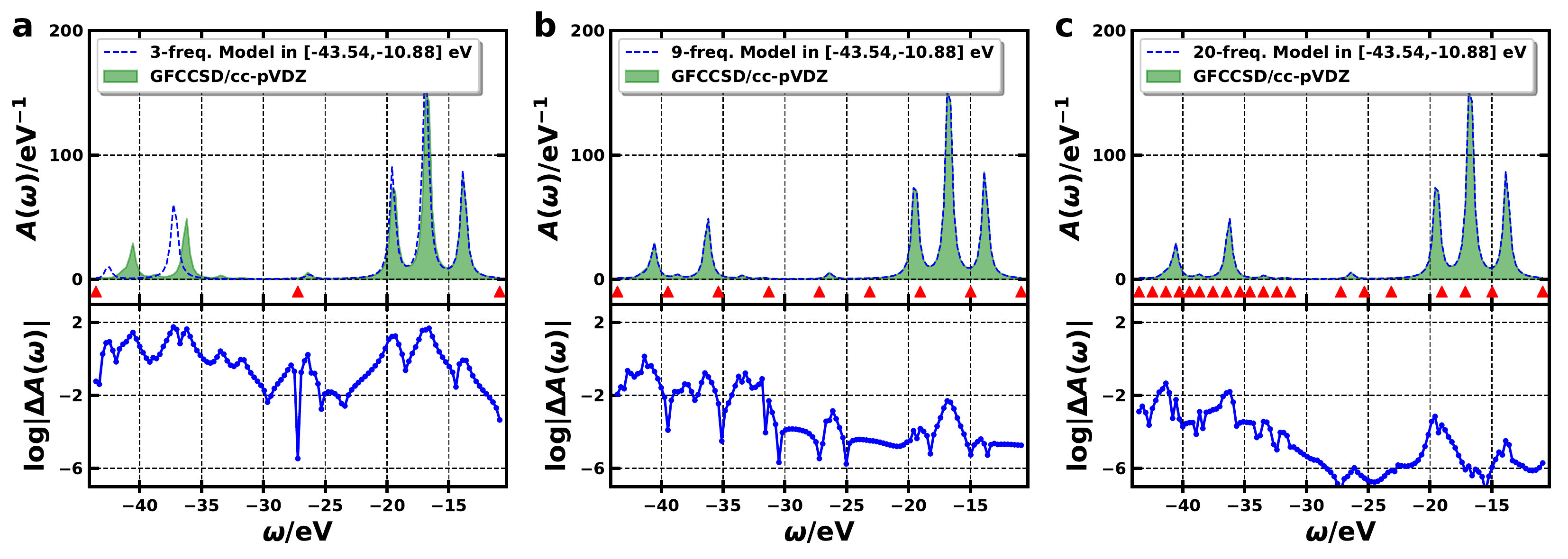}
\caption{Spectral functions of the carbon monoxide molecule in the energy regime of [-43.54, -10.88] eV
directly computed by conventional GFCCSD method (green shadow) and interpolated by its reduced order models (blue dashed line)
with $\eta$=0.027 eV and $\Delta\omega \le$0.027 eV The logarithmic values of the absolute difference between the 
GFCCSD and GFCCSD-MOR results are shown in the 
lower panel in each subplot. The selected frequencies whose corresponding
auxiliary vectors, $\mathbf{X}_p$, are used to construct the reduced order models are indicated by red 
upper triangle markers in each subplot.
\label{demo1}}
\end{sidewaysfigure}


\section{results and discussion}

In the following, we discuss the interpolated and/or extrapolated results from our GFCC method employing model order reduction technique (GFCC-MOR). The corresponding accuracy and efficiency tests are focused on the computed spectral functions of the carbon monoxide, {\it s-trans-}1,3-butadiene, benzene, and adenine molecules, and their comparisons to those computed by the original GFCC method and to other theoretical and experimental results. 
We used an experimental geometry\cite{herzberg2013} for the carbon monoxide molecule, and the geometries of {\it s-trans-}1,3-butadiene, benzene, and adenine are obtained from Ref. \citenum{setten13_232}, \citenum{chan12_4013}, and \citenum{krylov10_12305}, respectively. For each molecular system, conventional ($\Lambda$-)CCSD calculations were performed by using the Tensor Contraction Engine (TCE) module implemented in NWChem suite of quantum chemical codes.\cite{nwchem} The converged $T$ and $\Lambda$ amplitudes, as well as the two-electron integral tensor, are then used to compute the auxiliary vectors, $\textbf{X}_p$, and the GFCC matrix elements (and thereby the spectral functions). 

Fig. \ref{demo1} demonstrates how the expansion of the subspace improves the interpolated spectral function computed by the GFCCSD-MOR method for the carbon monoxide molecule. The calculation starts by solving the full-dimensional GFCC linear equations for the auxiliary vectors, $\textbf{X}_p$, at three frequencies evenly distributed over [-43.54, -10.88] eV. The converged $\textbf{X}_p$'s are then orthonormalized to construct the subspace $\textbf{S}$, upon which a reduced order model (denoted as a three-frequency model) is built via the manner depicted in Eq. (\ref{model}). The blue dashed line in Fig. \ref{demo1}a shows the interpolated spectral function obtained from the three-frequency model, in which the reduced order model is solved for every single frequency point on a discretization of [-43.54, -10.88] eV ($\Delta\omega\le$0.027 eV). The corresponding peak positions and the associated leading configurations at the peak positions are given in Tab. \ref{tab1}. Apparently, the interpolated results obtained from the three-frequency model is not able to reproduce the GFCCSD spectral function to a sufficient accuracy. In the regime of [-25, -10] eV, even though the positions of three main peaks roughly fit the full dimensional GFCCSD results, the average deviation of the spectral function in this regime with respect to the original GFCCSD result is above 0.1 eV$^{-1}$. For the regime below -25 eV, in comparison to the GFCCSD results, two satellites in [-35.00, -30.00] eV are missing from the result of the three-frequency model. For the remaining peaks, larger offsets and different leading configurations with respect to the original GFCCSD results can be observed (see Fig. \ref{demo1} and Tab. \ref{tab1}). For example, a $>$1 eV shift can be observed for the peak below -40 eV, and its associated leading configurations includes more contributions from $5\sigma^{-2}1\sigma^\ast$ and $1\pi^{-1}5\sigma^{-1}1\sigma^\ast$ two-electron processes which were barely found for the peaks in this regime in the original GFCCSD calculation.\cite{kowalski18_4335}

The reduced order model may be improved by expanding the span of the subspace $\textbf{S}$, i.e. including more orthonormalized $\textbf{X}_p$'s converged at some other frequencies picked by the aforementioned adaptive refinement strategy. As can be seen from Fig. \ref{demo1}b and Tab. \ref{tab1}, a nine-frequency model already shows a sizable improvement in comparison with the three-frequency model. In terms of the number of peaks, peak positions, and leading configurations relating to the peak, the nine-frequency model in Fig. \ref{demo1}b brings back the two missing peaks in [-35.00, -30.00] eV mentioned above, and produces negligible shifts and almost the same leading configurations for all the peaks within [-43.54, -10.88] eV with respect to original GFCCSD result. As for the peak height (i.e. the magnitude of the spectral function), for the outer valence regime ($>$25 eV) and the inner valence regime ($<$25 eV), the average deviations of the interpolated spectral function obtained from the nine-frequency model with respect to the original GFCCSD result are below 0.01 eV$^{-1}$ and 0.1 eV$^{-1}$, respectively. Of course, one can continue
to refine the aforementioned adaptive refinement strategy to further improve the model. For example, a 20-frequency model (Fig. \ref{demo1}c and Tab. \ref{tab1}) with more frequencies in the inner valence regime being picked produces same peak number, and almost same peak positions and their associated leading configurations, but significantly suppresses the maximum deviation of the spectral function in the inner valence regime from $\sim$1 eV$^{-1}$ (at $\sim$-41 eV in Fig. \ref{demo1}b) towards 0.01 eV$^{-1}$.

\begin{sidewaysfigure}
\includegraphics[width=\textwidth]{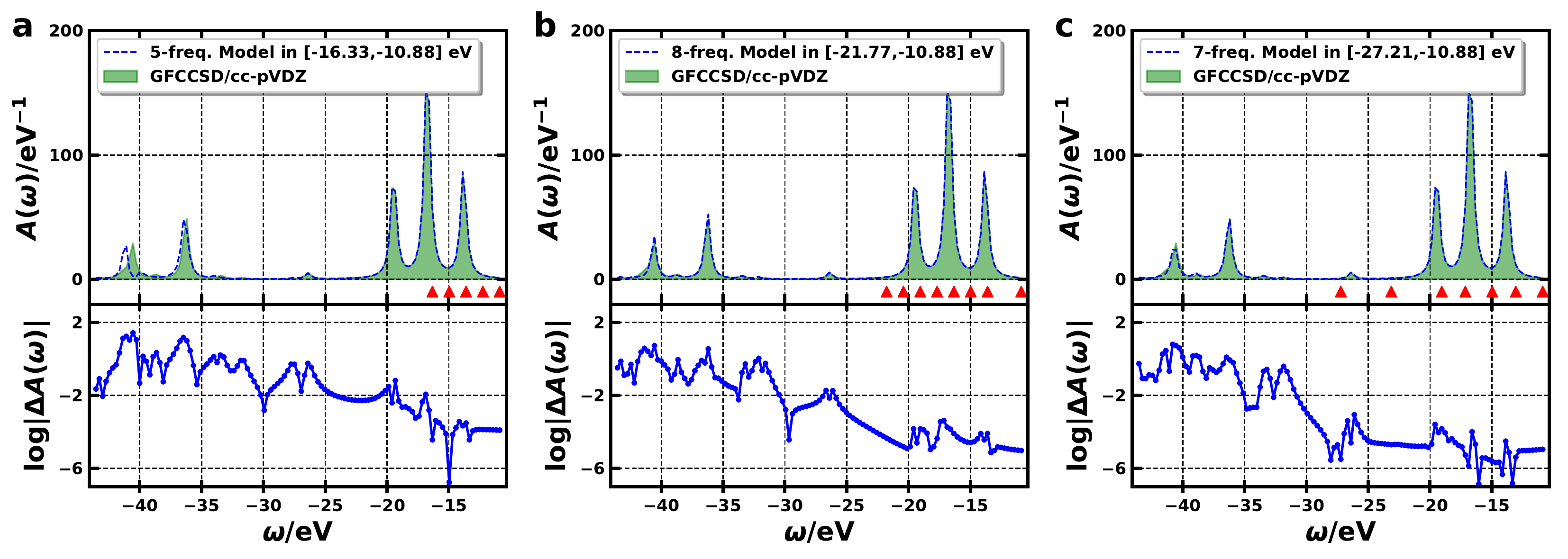}
\caption{Spectral functions of the carbon monoxide molecule in the frequency regime of [-43.54, -10.88] eV
directly computed by conventional GFCCSD method (green shadow) and interpolated and extrapolated by 
its reduced order models (blue dashed line)
with $\eta$=0.027 eV and $\Delta\omega \le$0.027 eV. The logarithmic values of the absolute difference between the 
GFCCSD and GFCCSD-MOR resutls are shown in the 
lower panel in each subplot. The selected frequencies whose corresponding
auxiliary vectors, $\mathbf{X}_p$, are used to construct the reduced order models are indicated by red 
upper triangle markers in each subplot.
\label{demo2}}
\end{sidewaysfigure}

In addition to interpolation, the reduced order model can also be utilized to produces extrapolated results over an extended frequency window. To assess the accuracy of this model extrapolation, some demonstrative tests have been performed using the carbon monoxide system previously
discussed. The results of this assessment are outlined in Fig. \ref{demo2}  and Tab. \ref{tab1}. In this example, approximate spectral functions were produced on the domain [-43.54, -10.88] eV
using three different reduced order models built from different subsets of the original frequency regime (namely, [-16.33, -10.88] eV, [-21.77, -10.88] eV, and [-27.21,-10.88] eV), and are compared with the original GFCCSD spectral function.
As is apparent, two of the three models (excluding the five-frequency model in [-16.33, -10.88] eV) produce the same number of peaks as the GFCCSD results. In the result from the five-frequency model in [-16.33, -10.88] eV, a low magnitude satellite is missing at $\sim$-32 eV. As for peak positions and the associated leading configurations, except for the missing peak just discussed, all three models give consistent results with the original GFCCSD results. The largest deviation ($\sim$0.55 eV) occurs for the satellite peak (at -39.27 eV) obtained from the extrapolation of the five-frequency model in [-16.33, -10.88] eV. Further, the leading configurations of that peak also show an increased contribution from $3\sigma^{-1}$ ionization process which was found to be trivial in the original GFCCSD calculation.\cite{kowalski18_4335} Nevertheless, when the model covers wider frequency regime (as can be seen from Fig. \ref{demo2}b,c and Tab. \ref{tab1}), the extrapolated spectral functions give almost the same peak positions and leading configurations as the original GFCCSD results, and the deviation of the corresponding spectral functions also gradually decreases. 

Given the more or less same number of frequencies (and thereby the number of $\textbf{X}_p$'s for constructing the subspace), it is instructive to compare the models shown in Figs. \ref{demo1}b and \ref{demo2}b. Both models successfully reproduce the original GFCCSD peak multiplicity and positions, and qualitatively reproduce the spectral function in the studied frequency regime. However, there is a major difference between the two models in terms of the frequency points distribution, i.e. the model in Fig. \ref{demo1}b requires five frequency points in the inner valence regime of [-45.00, -25.00] eV, while all the frequency points required by the model in Fig. \ref{demo2}b are located in the outer valence regime ($>$-25 eV). Note that, due to stronger many-body effect in the inner valence regime, it usually requires much more effort for the linear solver to converge the GFCC linear system for the $\textbf{X}_p$'s in that domain than it does for the $\textbf{X}_p$'s in the outer valence regime.\cite{kowalski18_4335, kowalski18_214102} 
Therefore, in comparison to the model in Fig. \ref{demo1}b, the model in Fig. \ref{demo2}b in general requires much less effort to be built, while still being able to achieve the same level of accuracy for the studied frequency regime. 

\begin{table*}
\caption{Vertical ionization potentials of the CO molecule, as well as associated molecular orbital ($p$) and leading configurations (with associated contribution $>$0.1), obtained from the GFCCSD-MOR calculations shown in Figs. \ref{demo1} and \ref{demo2}. The cc-pVDZ basis is used in all the calculations. The unit of all frequencies is eV.}
\label{tab1}
\resizebox{\columnwidth}{!}{
\begin{tabular}{c l c l l l c l c l l l c l c l l}
\\
\hline
\multicolumn{5}{c}{3-freq. model in [-43.54, -10.88]} &&
\multicolumn{5}{c}{9-freq. model in [-43.54, -10.88]} &&
\multicolumn{5}{c}{20-freq. model in [-43.54, -10.88]}
\\
\cline{1-5} \cline{7-11} \cline{13-17} 
$\omega$ && $p$ && Leading Conf. &&
$\omega$ && $p$ && Leading Conf. &&
$\omega$ && $p$ && Leading Conf. 
\\
\cline{1-17}
-13.825 && 7 && $5\sigma^{-1}$
&& 
-13.797 && 7 && $5\sigma^{-1}$
&&
-13.797 && 7 && $5\sigma^{-1}$
\\
-16.873 && 5,6 && $1\pi^{-1}$
&&
-16.737 && 5,6 && $1\pi^{-1}$
&& 
-16.737 && 5,6 && $1\pi^{-1}$
\\
-19.512 && 4 && $4\sigma^{-1}$
&& 
-19.458 && 4 && $4\sigma^{-1}$
&&
-19.458 && 4 && $4\sigma^{-1}$
\\
-26.289 && 3,4,7 && $5\sigma^{-1}1\pi^{-1}1\pi^\ast$
&& 
-26.370 && 3,4,7 && $5\sigma^{-1}1\pi^{-1}1\pi^\ast$
&&
-26.370 && 3,4,7 && $5\sigma^{-1}1\pi^{-1}1\pi^\ast$
\\
$-$ 	     && $-$ && $-$
&& 
-31.759 && 5,6 && $5\sigma^{-2}1\pi^\ast$
&&
-31.704 && 5,6 && $5\sigma^{-2}1\pi^\ast$
\\
$-$        && $-$ && $-$
&& 
-33.446 && 3,4,7 && $5\sigma^{-1}1\pi^{-1}1\pi^\ast$
&&
-33.419 && 3,4,7 && $5\sigma^{-1}1\pi^{-1}1\pi^\ast$
\\
-36.439 && 3 && $3\sigma^{-1}$
&& 
-36.249 && 3 && $3\sigma^{-1}$
&&
-36.249 && 3 && $3\sigma^{-1}$
\\
-40.685 && 3 && $3\sigma^{-1}$, $5\sigma^{-2}1\sigma^\ast$
&& 
-38.725 && 3 && $5\sigma^{-2}1\sigma^\ast$
&&
-38.725 && 3 && $5\sigma^{-2}1\sigma^\ast$
\\
-42.535 && 3 && $5\sigma^{-2}1\sigma^\ast$, $1\pi^{-1}4\sigma^{-1}1\pi^\ast$,
&& 
-40.576 && 3 && $3\sigma^{-1}$, $1\pi^{-1}4\sigma^{-1}1\pi^\ast$
&&
-40.576 && 3 && $3\sigma^{-1}$, $1\pi^{-1}4\sigma^{-1}1\pi^\ast$
\\
&& && $1\pi^{-1}5\sigma^{-1}1\pi^\ast$, $3\sigma^{-1}$
&& 
&& &&
&&
&& &&
\\
\hline
\multicolumn{5}{c}{5-freq. model in [-16.33, -10.88]} &&
\multicolumn{5}{c}{8-freq. model in [-21.77, -10.88]} &&
\multicolumn{5}{c}{7-freq. model in [-27.21, -10.88]}
\\
\cline{1-5} \cline{7-11} \cline{13-17} 
$\omega$ && $p$ && Leading Conf. &&
$\omega$ && $p$ && Leading Conf. &&
$\omega$ && $p$ && Leading Conf. 
\\
\cline{1-17}
-13.797 && 7 && $5\sigma^{-1}$
&& 
-13.797 && 7 && $5\sigma^{-1}$
&&
-13.797 && 7 && $5\sigma^{-1}$
\\
-16.737 && 5,6 && $1\pi^{-1}$
&&
-16.737 && 5,6 && $1\pi^{-1}$
&& 
-16.737 && 5,6 && $1\pi^{-1}$
\\
-19.458 && 4 && $4\sigma^{-1}$
&& 
-19.458 && 4 && $4\sigma^{-1}$
&&
-19.458 && 4 && $4\sigma^{-1}$
\\
-26.397 && 3,4,7 && $5\sigma^{-1}1\pi^{-1}1\pi^\ast$
&& 
-26.370 && 3,4,7 && $5\sigma^{-1}1\pi^{-1}1\pi^\ast$
&&
-26.370 && 3,4,7 && $5\sigma^{-1}1\pi^{-1}1\pi^\ast$
\\
$-$ 	     && $-$ && $-$
&& 
-31.895 && 5,6 && $5\sigma^{-2}1\pi^\ast$
&&
-31.759 && 5,6 && $5\sigma^{-2}1\pi^\ast$
\\
-33.800 && 3,4,7 && $5\sigma^{-1}1\pi^{-1}1\pi^\ast$
&& 
-33.419 && 3,4,7 && $5\sigma^{-1}1\pi^{-1}1\pi^\ast$
&&
-33.446 && 3,4,7 && $5\sigma^{-1}1\pi^{-1}1\pi^\ast$
\\
-36.331 && 3 && $3\sigma^{-1}$
&& 
-36.249 && 3 && $3\sigma^{-1}$
&&
-36.249 && 3 && $3\sigma^{-1}$
\\
-39.270 && 3 && $5\sigma^{-2}1\sigma^\ast$, $3\sigma^{-1}$
&& 
-38.834 && 3 && $5\sigma^{-2}1\sigma^\ast$
&&
-38.853 && 3 && $5\sigma^{-2}1\sigma^\ast$
\\
-40.875 && 3 && $3\sigma^{-1}$, $1\pi^{-1}4\sigma^{-1}1\pi^\ast$
&& 
-40.630 && 3 && $3\sigma^{-1}$, $1\pi^{-1}4\sigma^{-1}1\pi^\ast$
&&
-40.651 && 3 && $3\sigma^{-1}$, $1\pi^{-1}4\sigma^{-1}1\pi^\ast$
\\
\end{tabular}
}
\end{table*}

\begin{figure}
\includegraphics[width=0.5\textwidth]{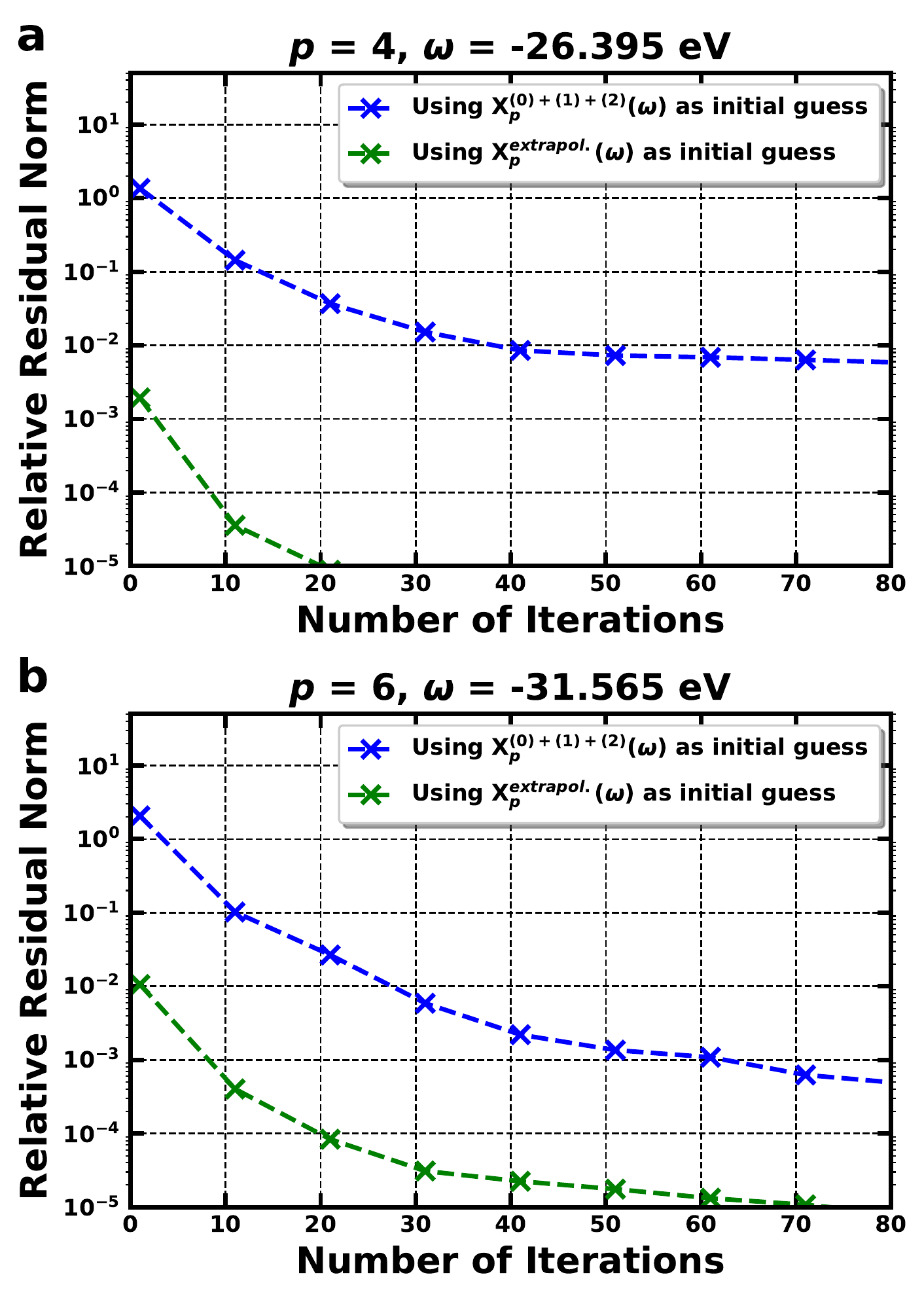}
\caption{Convergence performance test of the linear equation solver (direct inversion in the iterative subspace, or DIIS, adapted to complex space) employing different initial guess vectors for the  calculations of ${\bf X}_p$'s at chosen orbital indices, $p$'s, and frequencies, $\omega$'s, for the carbon monoxide molecule. The number of micro-iteration iterates used in the test is 10. ({\bf a}) corresponds to the GFCCSD calculations of the ${\bf X}_p$'s significantly contributing to the first satellite peak in the valence band of the spectral function. ({\bf b}) corresponds to the GFCCSD calculations of the ${\bf X}_p$'s significantly contributing to the second satellite peak in the valence band of the spectral function. $\mathbf{X}_p^{(0)+(1)+(2)}$ is the sum of 0th, 1st, and 2nd order terms of the $\mathbf{X}_p$ that have been given in Ref. \citenum{kowalski16_062512}. $\mathbf{X}_p^{extrapol.}$ is obtained by extrapolating the eight-frequency model in [-21.77, -10.88] eV shown in Fig. \ref{demo2}b to the designated frequencies.
\label{conv}}
\end{figure}

Remarkably, if higher accuracy for the extended frequency regime is desired, the results given by extrapolating the model from a narrower frequency window may still be useful as they can often be used as an effective initial guess to aid in the convergence of the iterative linear solver used to solve
the GFCC equations. 
Further, as the reduced
order linear systems may be solved directly, the use of these approximate solutions as initial
guesses for the full dimensional iterative solver can be of particular use in cases where
convergence of these solutions proves difficult.
Typical examples of this state of affairs are shown in Fig. \ref{conv}, where convergence profiles (i.e. the residual norm of the linear equation going along with the iterations) resulting from using different initial guesses in the same iterative linear solver are compared at two frequencies in the inner valence regime (namely at -31.565 eV and -26.395 eV, respectively). In this regime, there are two satellites dominated by the $5\sigma^{-2}1\pi^\ast$ and $5\sigma^{-1}1\pi^{-1}1\pi^\ast$ two-electron processes, respectively. In the blue curves depicted in Fig. \ref{conv}, the linear solver employs up to the second order perturbative terms of the $\textbf{X}_p$'s as the initial guess of the solution (the analytic perturbative terms of the $\textbf{X}_p$'s were given in Ref. \citenum{kowalski16_062512}), while in the green curve the linear solver employs the extrapolated results (from the eight-frequency model in Fig. \ref{demo2}b) as the initial guess of the solution.
As can be seen, in comparison with using the low-order perturbative terms of the $\textbf{X}_p$'s, using the extrapolated results as the initial guess for the
iterative solver can drastically improve the convergence profile for the solutions of interest. For example, after the first iteration, the relative residual norms shown in the green curves are at least two orders of magnitude smaller than the ones in the blue curves for both cases. As iterations progress, the blue curves fail to converge (i.e. relative residual norm being $<10^{-5}$) within 80 steps in both case (indicating that a larger number of micro-iterations iterates and more iterations may be needed in the DIIS linear solver used in this study). In contrast, using the MOR extrapolated results as an initial guess for the linear solver convergence is achieved within 80 iterations in both of the cases examined. It is worth mentioning that the convergence would generally be expected to be improved if higher order terms are included in the initial guess. However, this can be time-consuming in practice since the computational complexity of computing the third and higher order terms of $\textbf{X}_p$'s\cite{kowalski16_062512} would become similar to that of solving the GFCC linear equation itself.

\begin{figure}
\includegraphics[width=1.0\textwidth]{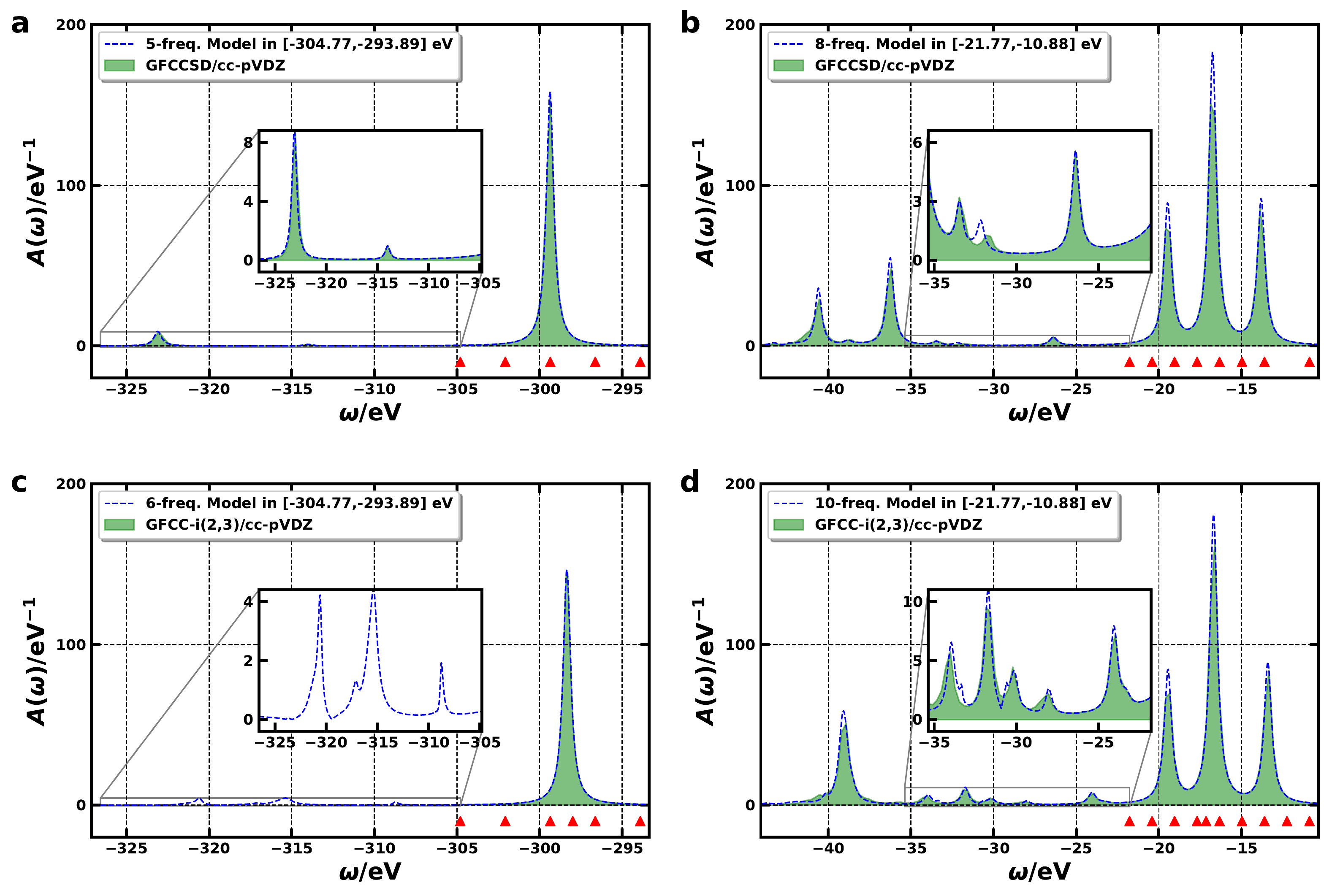}
\caption{Spectral functions of the carbon monoxide molecule in the core and valence energy regimes directly computed by conventional GFCCSD and GFCC-i(2,3) methods (green shadow), and the corresponding GFCC-MOR methods (blue dashed line) with $\eta$=0.027 eV and $\Delta\omega \le$0.027 eV. The selected frequencies whose corresponding auxiliary vectors, $\mathbf{X}_{p}$, are used to construct the reduced order models are pointed out by red upper triangle markers in each subplot. The GFCC results are obtained from Refs. \citenum{kowalski18_4335} and \citenum{kowalski18_214102}. The insets show the magnified the satellite peaks in the designated regimes.
\label{demo3}}
\end{figure}

In the following, we examine the interpolative and extrapolative capabilities of
the proposed GFCC-MOR method in differing frequency regimes than the ones previously
discussed and in relation to different GFCC methods.
Fig. \ref{demo3} shows the spectral functions of the carbon monoxide molecule in both core and valence regimes computed by the conventional GFCC methods (including GFCCSD and GFCC-i(2,3)\cite{kowalski18_214102}) and the corresponding GFCC-MOR methods. Note that in comparison with the GFCCSD method, the GFCC-i(2,3) method includes the treatment of inner triples, which was demonstrated in our previous work being able to improve the many-body description of both main ionic states and satellites. In the core regime, the reduced order models are constructed in the regime of [-304.77, -293.89] eV, and the resulting interpolatory and extrapolatory features are examined in the broader regime of [-326.57, -293.91] eV. In the valence regime, the models are constructed in the regime of [-21.77, -10.88] eV, and interpolated and extrapolated to the broader regime of [-43.54, -10.88] eV. 
As can be seen, the profiles of the spectral functions obtained from the GFCC-MOR calculations approximate the available original GFCC results relatively well in both core and valence regimes. Tiny differences are only detected in the middle valence regime, where, in comparison to the GFCCSD results, the eight-frequency model in [-21.77, -10.88] eV blue-shifts the satellite located at $\sim$-32.00 eV by $\sim$0.55 eV (see the inset of Fig. \ref{demo3}b), and in comparison to the GFCC-i(2,3) results, the result from the ten-frequency model in [-21.77, -10.88] eV exhibits some tiny shoulder peaks located in [-35.00, -30.00] eV (see the inset of Fig. \ref{demo3}d). 

\begin{figure}
\includegraphics[width=0.5\textwidth]{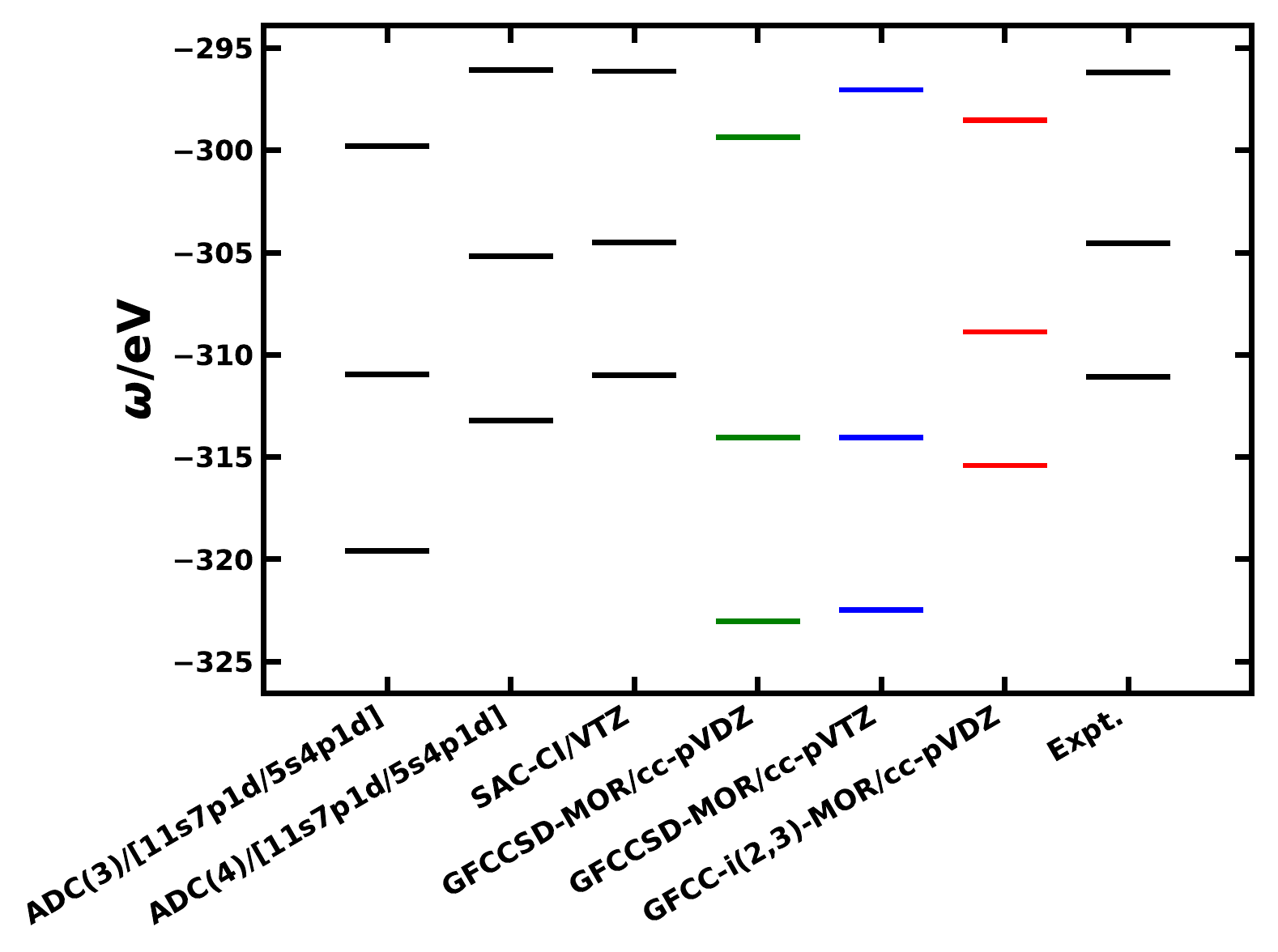}
\caption{Comparison of different theoretical ionization potentials and experimental values. The GFCC-MOR results are read from Fig. \ref{demo3}a,c. The experimental values, SAC-CI, and ADC results are collected from Refs. \citenum{ehara06_114304} and \citenum{schirmer87_6031}. 
\label{core_test}}
\end{figure}
 
Note that, in Fig. \ref{demo3}c (and its inset), by extrapolating the six-frequency model from the regime of [-304.77, -293.89] eV to [-326.57, -293.89] eV, three significant satellites, which were not reported in the previous GFCC-i(2,3) calculations,\cite{kowalski18_214102} are identified at -320.85, -315.41, and -308.88 eV, respectively (a weaker shoulder peak can also be seen at $\sim$-316.77 eV). The configuration analysis shows that these satellites are mainly attributed to the $1\pi^{-1}2\sigma^{-1}1\pi^\ast$ two-electron process (classified as the so-called direct shake-up states), and are consistent with the configuration analysis of the satellites obtained from GFCCSD/GFCCSD-MOR calculations in the regime of [-326.57, -293.91] eV (see the inset of Fig. \ref{demo3}a), as well as other theoretical analysis.\cite{ehara06_114304, schirmer87_6031} To further confirm the accuracy of the core level results obtained from the GFCC-MOR calculations, we compare the positions of the main peak and two rightmost satellites observed in Fig. \ref{demo3}a,c with the available ionization potentials computed from other theories, as well as the corresponding experimental values. 
As can be seen from Fig. \ref{core_test}, similar to the difference between ADC(4) and ADC(3) results, since higher-order excitations (e.g. ionic triple excitations) are taken into account in the schemes,\cite{kowalski18_214102} the GFCC-i(2,3)-MOR results show significant improvement over the GFCCSD-MOR results, especially the positions of the two satellites are red-shifted by $\sim$5.0 eV and $\sim$8.0 eV, respectively.
Even though the GFCC-i(2,3)-MOR results still underestimate the experimental values by 1$\sim$2 eV, the relative positions of the three peaks are very close to the experimental values and SAC-CI results. In particular, the spacing between the two satellites ($\sim$6.5 eV) in the GFCC-i(2,3)-MOR results is almost same as what are observed from the experiment and the SAC-CI calculation, and is better than the ADC(4) results ($\sim$8.0 eV).

\begin{figure}
\includegraphics[width=\textwidth]{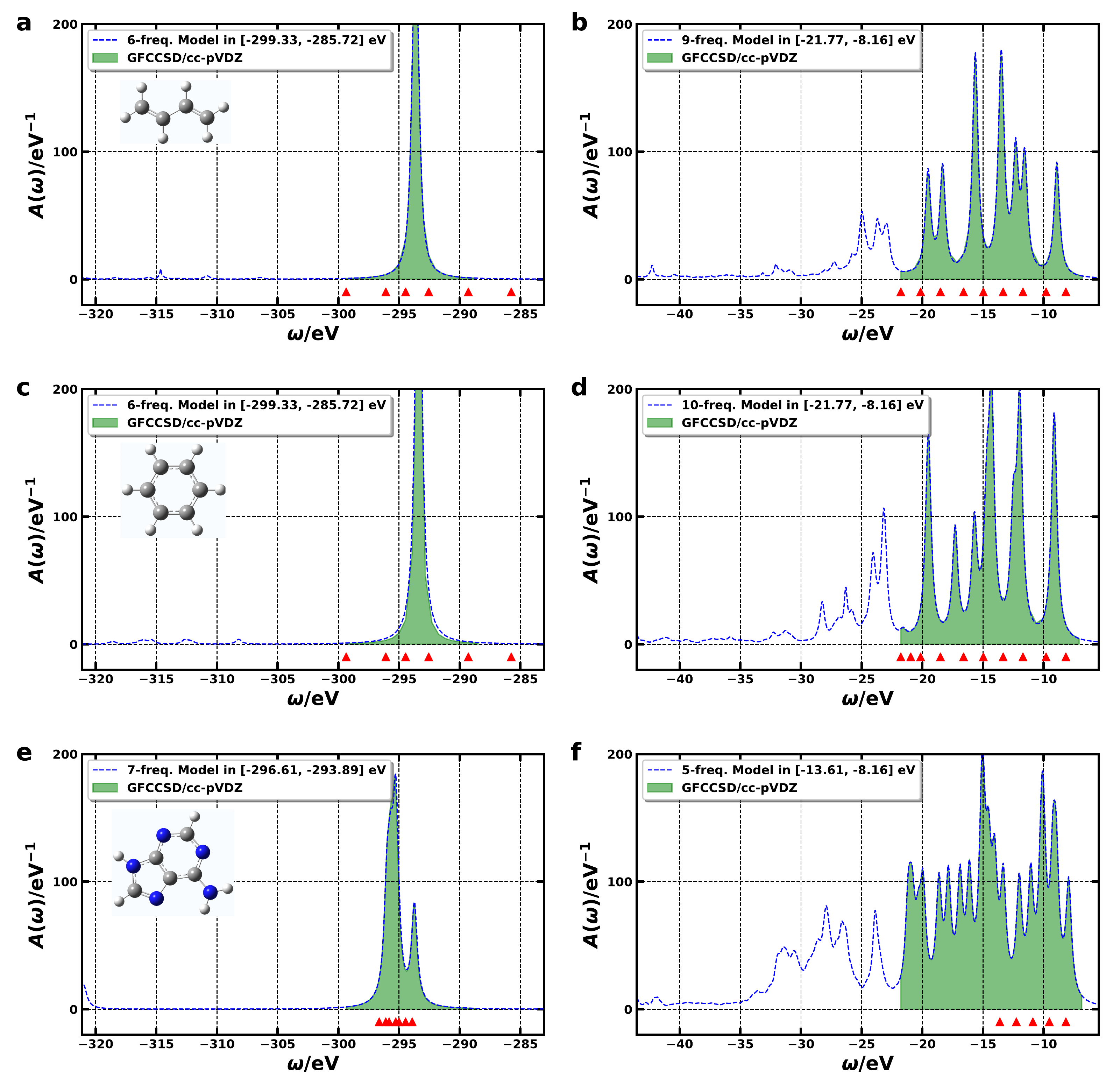}
\caption{Spectral functions of the 1,3-butadiene, benzene, and adenine molecules in the core and valence energy regimes
directly computed by conventional GFCCSD method (green shadow), and interpolated and extrapolated by 
their reduced order models (blue dashed line)
with $\eta$=0.027 eV and $\Delta\omega \le$0.027 eV. The selected frequencies whose corresponding
auxiliary vectors, $\mathbf{X}_{p}$, are used to construct the reduced order models are indicated by red 
upper triangle markers in each subplot.
\label{demo4}}
\end{figure}

\begin{figure}
\includegraphics[width=0.5\textwidth]{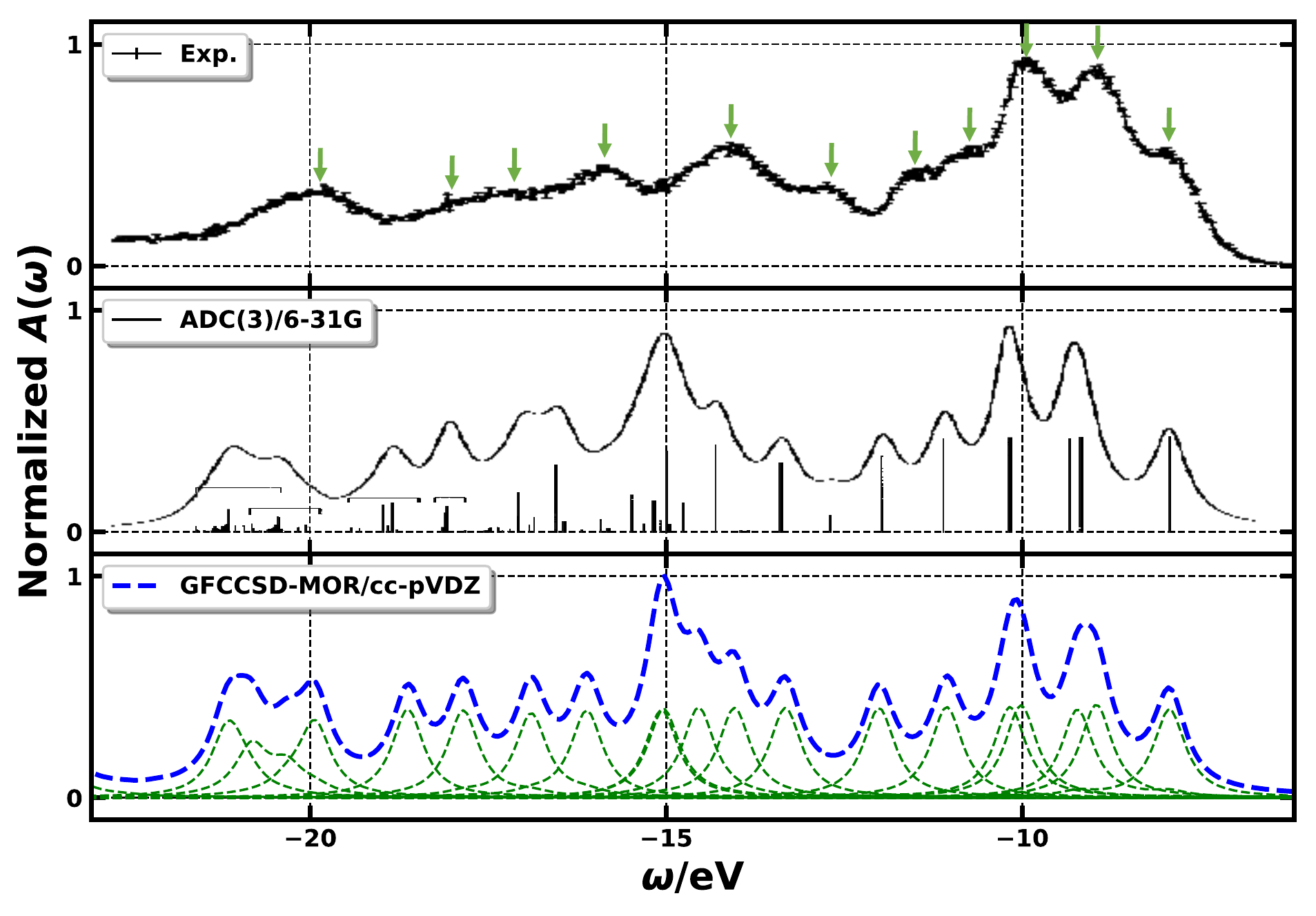}
\caption{The spectral function of the adenine molecule in the valence energy regimes
directly computed by the GFCCSD-MOR method (green shadow), and in comparison with previously reported ADC(3) and experimental results. The experimental PES was recorded at a photon energy of 80 eV for $\theta$=0$^o$, and peak positions are labeled by green arrows. The eigenvalues and the associated pole strength from the ADC(3) calculation are labeled by bars. The ADC(3) results and the PES spectrum are collected from Ref. \citenum{trofimov06_305}.
\label{ade}}
\end{figure}

To further test the efficiency of the proposed GFCC-MOR method, we examine its
application to some larger systems. In Fig. \ref{demo4}, the spectral functions of the 1,3-butadiene, benzene, and adenine molecules over broad core and valence energy regimes computed by the GFCCSD-MOR methods (blue dashed line) are exhibited and compared with the reported GFCCSD results (green shadow). The reported GFCCSD results are shown between -300.00 eV and -285.00 eV for the core regime (Fig. \ref{demo4}a,c,e), and between -19.00 eV and -7.00 eV for the valence regime (Fig. \ref{demo4}b,d,f). For 1,3-butadiene and benzene, the auxiliary vectors used for the model construction are collected from the frequency window of [-299.33, -285.72] eV in the core regime and the frequency window of [-21.77, -8.16] eV in the valence regime, respectively. For the larger system, the adenine molecule, the auxiliary vectors are collected from the frequency windows of [-296.61, -293.89] eV in the core regime and [-13.61, -8.16] eV in the valence regime, respectively, due to the fact that more peak features are exposed in these narrower frequency windows. 
As can be seen, the spectral functions obtained from the combination of the interpolated and extrapolated results of the reduced order model fit the original GFCCSD results pretty well in the designated frequency windows spread over both core and valence regimes. In particular, the five-frequency model from the frequency window of [-13.61, -8.16] eV is able to well reproduce the original GFCCSD peak features between -21.77 eV and -8.00 eV. 
We further compare the reduced order model results with available theoretical results from other Green's function methods and experimental photoelectron spectra (PES). For example, in Fig. \ref{ade}, the extrapolated spectral function of adenine molecule in this work is compared with the reported ADC result and the photoelectron spectrum (PES). As can be seen, in comparison to the ADC(3) results, the GFCCSD-MOR calculation gives almost same results in terms of peak positions and heights. In comparison with the experimental spectrum, the GFCCSD-MOR spectral function gives almost same peak positions and peak profile as the PES in [-12.00, -4.00] eV. 

As we discussed in our previous work,\cite{kowalski18_214102} special attention should be paid when comparing the spectral function profile with the PES results. In principle, the spectral function (or the pole strength) is computed through the overlap integrals between the initial states and their final ionic states. The relative peak intensities in the PES requires not only the calculation of the overlap integrals, but also the dipole integrals describing the dipole interactions between the initial states and their corresponding final ionic state manifolds. Therefore, the relative intensities exhibited in the PES are expected to be different from those observed from the calculated spectral functions, except that when the photon energy is much higher than the ionization potentials in the PES measurement, and there are no appreciable initial state correlation effects for the system. From this perspective, the reason why the GFCCSD-MOR spectral function gives almost same peak positions and peak profile as the PES in [-12.00, -4.00] eV can be attributed to the following observations, (i) the photon energy used to record the PES (i.e. 80 eV) is much higher than the energy regime, and (ii) the effects of electron correlation in this energy regime are weak such that even the results from the single-particle theory (e.g. outer-valence Green's function method) fits the PES very well (see Fig. 10 in Ref. \citenum{trofimov06_305}). When going beyond this regime, since the aforementioned two conditions are gradually violated as witnessed by the appearance of more and more satellite states that requires the treatment of many-body effects, the comparison between the computed spectral function and the PES should only be in the peak positions. Here, for the regime below -12 eV, both the ADC(3) and GFCCSD-MOR peak positions are systematically blue-shifted by $\sim$1 eV with respect to the PES peaks, but the relative positions of these peaks still fit the PES very well.

\begin{figure}
\includegraphics[width=0.5\textwidth]{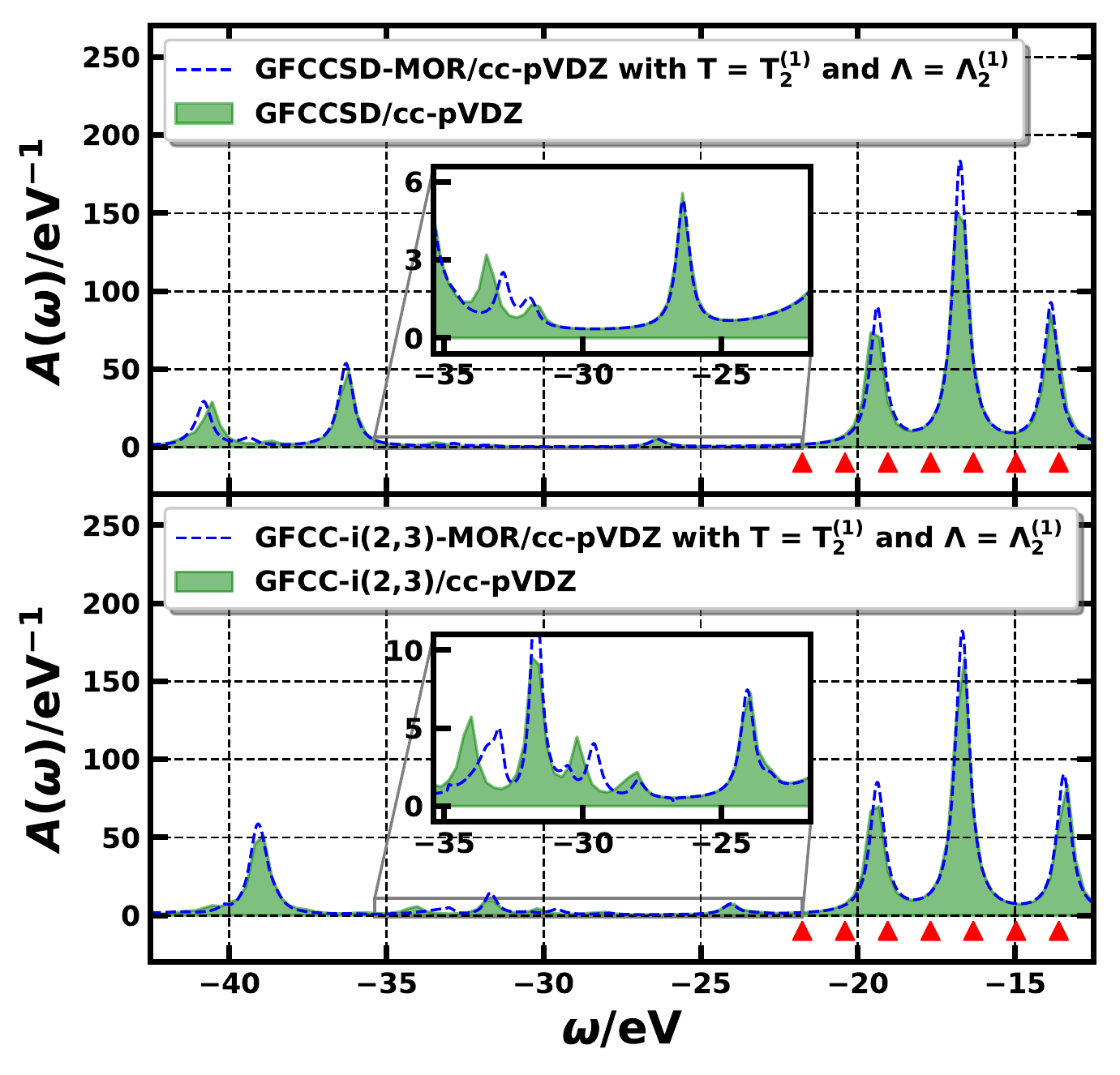}
\caption{The spectral functions of the carbon monoxide molecule in the valence energy regimes computed by the GFCC-MOR methods with amplitudes $T$ and $\Lambda$ being replaced by their first order perturbative terms. The selected frequencies whose corresponding auxiliary vectors, $\textbf{X}_p$, are used to construct the reduced order models are indicated by red upper triangle markers.
\label{pert}}
\end{figure}

It is important to emphasize that while the conventional GFCC and proposed
GFCC-MOR methods presented in this work yield similar quality results, the latter is
significantly more efficient than the former.
For the spectral function calculation in the frequency regime of [-43.54, -10.88] eV, the conventional GFCC calculations requires the solution of the full dimensional GFCC linear equation at every single frequency point in a 1200 points discretization (if $\Delta\omega\le$0.027 eV), while the GFCC-MOR calculations only requires the solution of the full dimensional GFCC linear equation at 5$\sim$10 frequency points. 
Due to this large disparity between the number
of the solutions of the full dimensional linear equation required between the two methods, the computational
cost of constructing the reduced order model from the small set of solutions required by
the GFCC-MOR method is negligible in contrast to the cost of solving the linear systems
themselves. As such, the overall computational cost of the GFCC-MOR
method presented in this work is roughly two orders of magnitude less than that of the
conventional GFCC method.


\section{Conclusion and outlook}

In this work, we
demonstrated the efficacy of the application of model order reduction to the GFCC method
(the GFCC-MOR method) in efforts to reduce the computational cost of evaluating the CC
Green's function in specific spectral regimes.
%
In the conventional GFCC method,
one must solve the full dimensional GFCC linear system at every frequency point in some
discretization of a spectral region of interest to evaluate the CC Green's function. In
contrast, the GFCC-MOR method reduces the computational cost of evaluating the CC
Green's function by selecting only a few points in the spectral region of interest to solve
the full dimensional linear systems, and then projects this problem onto a model basis of
reduced order which is constructed from the full dimensional auxiliary vectors at these
frequencies. The dimension reduction in the reduced order model allows for rapid evaluation
of an approximate CC Green's function in a (possibly fine) discretization of the spectral
region of interest.
Relevant accuracy comparisons have been performed for computing the spectral functions of the carbon monoxide, 1,3-butadiene, benzene, and adenine molecules in both the core and valence regimes. The GFCC-MOR results show excellent agreement with the full dimensional GFCC results. By extrapolation, the GFCC-MOR method may also be utilized to either predict the results in some energy regimes where the original GFCC results are unavailable, or provide a reasonable initial guess to aid in the convergence of the full dimensional GFCC linear equation in that regime.  Moreover, the GFCC-MOR method has been shown to be very efficient, and  the associated cost of approximating the GFCC spectral functions for the molecules studied in this work was found to be roughly two orders of magnitude smaller than the original GFCC method. Therefore, the MOR technique has shown a great capability to offload a large computational overhead from the original GFCC method, and a great potential to broaden the large-scale applications of this method.

To further reduce the computational cost of the GFCC-MOR method, the MOR technique may be applied to the GFCC method in a different way. For example, one can build separate reduced order models for every single diagonal Green's function curve directly (and the sum of these curves then gives the approximate spectral function). Besides, one can also consider employing the perturbative cluster amplitudes in the GFCC-MOR method. For example, as discussed in the earlier GFCC work of Nooijen and Snijders,\cite{nooijen95_1681}, the cluster amplitudes used in the GFCC-MOR method may be further replaced by their first-order perturbative terms. Apparently, this approximation further accelerate the calculations by skipping the expensive CC ground state calculations, and does not affect the connectedness of the GFCC diagrams.  As can be seen from some preliminary results for the carbon monoxide (see Fig. \ref{pert}), even though some small deviations ($<$1 eV) for the satellites are observed in the inner valence regime, this approximation works relatively well for reproducing the entire spectral function profile over the designated regime computed by the conventional GFCC method. The subspace construction and utilization in these directions are currently under intensive development, and more tests and discussions will be presented in our future papers.


\section{acknowledgement}

This work was supported by the Center for Scalable, Predictive methods for Excitation and Correlated phenomena (SPEC), which is funded by the U.S. Department of Energy (DOE), Office of Science, Office of Basic Energy Sciences, the Division of Chemical Sciences, Geosciences, and Biosciences.
All calculations have been performed using the Molecular Science Computing Facility (MSCF) in the Environmental Molecular Sciences Laboratory (EMSL) at the Pacific Northwest National Laboratory (PNNL). EMSL is funded by the Office of Biological and Environmental Research in the U.S. Department of Energy. PNNL is operated for the U.S. Department of Energy by the Battelle Memorial Institute under Contract DE-AC06-76RLO-1830. B. P. acknowledges the Linus Pauling Postdoctoral Fellowship from PNNL. 

\bibliography{gfcc-sub}
\newpage 

\begin{figure}
  \includegraphics[width=\columnwidth]{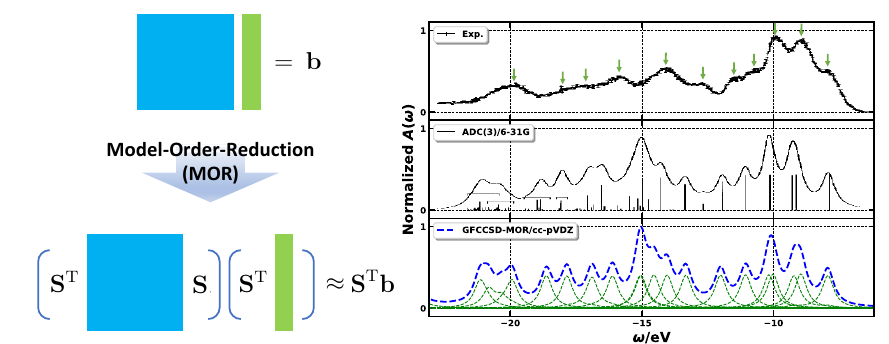}
\end{figure}

\end{document}